\documentclass[aps,prb,preprint,superscriptaddress]{revtex4-2}
\usepackage[T1,T2A]{fontenc}
\usepackage{bm, amssymb}
\usepackage{mathrsfs} 
\usepackage[utf8]{inputenc}
\usepackage{amsmath}
\usepackage{dcolumn}
\usepackage{graphicx}
\usepackage[dvips]{epsfig}
\usepackage{xcolor}

\usepackage[figuresright]{rotating}
\usepackage{bm}
\usepackage{amssymb}

\begin{document}

\title{Random magnetic anisotropy driven transitions in layered perovskite LaSrCoO$_4$}

\author{Abdul Ahad}
\thanks{abdul.ahad@ntu.edu.sg}
\thanks{Present Address: School of Physical and Mathematical Sciences, Nanyang Technological University, Singapore}
\address{Department of Physics, Aligarh Muslim University, Aligarh 202002, India}

\author{K. Gautam}
\thanks{Present Address:  RIKEN Center for Emergent Matter Science, Saitama, Japan}
\address{UGC-DAE Consortium for Scientific Research, Indore 452001, India}
\author{S. S. Majid}
\thanks{Present Address: National Institute of Technology Hazratbal Srinagar J \& K, India}
\address{Department of Physics, Aligarh Muslim University, Aligarh 202002, India}
\author{K. Dey}
\address{UGC-DAE Consortium for Scientific Research, Indore 452001, India}
\author{A. Tripathy}
\address{UGC-DAE Consortium for Scientific Research, Indore 452001, India}
\author{F. Rahman}
\address{Department of Physics, Aligarh Muslim University, Aligarh 202002, India}
\author{R. J. Choudhary}
\address{UGC-DAE Consortium for Scientific Research, Indore 452001, India}

\author{R. Sankar}
\address{Institute of Physics, Academia Sinica, Taipei 11529, Taiwan}
\author{A.K. Sinha}
\thanks{Present address: Department of Physics, School of Engineering, University of Petroleum and energy studies, Dehradun, India}
\address{Indus Synchrotrons Utilization Division, Raja Ramanna Center for Advanced Technology, Indore 452013, India}
\author{S. N. Kaul}
\thanks{sn.kaul@uohyd.ac.in}
\address{School of Physics, University of Hyderabad, Hyderabad 500046, India}
\author{D. K. Shukla}
\thanks{dkshukla@csr.res.in}
\address{UGC-DAE Consortium for Scientific Research, Indore 452001, India}

\begin{abstract}
Attempts to unravel the nature of magnetic ordering in LaSrCoO$_4$ (Co$^{3+}$), a compound intermediate between antiferromagnetic (AFM) La$_2$CoO$_4$ (Co$^{2+}$) and ferromagnetic (FM) Sr$_2$CoO$_4$ (Co$^{4+}$), have met with a limited success so far. In this report, the results of a thorough investigation of dc magnetization and ac susceptibility (ACS) in single-phase LaSrCoO$_4$ provide clinching evidence for a thermodynamic paramagnetic (PM) - ferromagnetic (FM) phase transition at T$_{c}$ = 220.5 K, followed at lower temperature (T$_{g}$ = 7.7 K) by a transition to the cluster spin glass (CSG) state. Analysis of the low-field Arrott plot isotherms, in the critical region near T$_{c}$, in terms of the Aharony-Pytte scaling equation of state clearly establishes that the PM-FM transition is basically driven by random magnetic anisotropy (RMA). For temperatures below $\approx$ 30 K, large enough RMA destroys long-range FM order by breaking up the infinite FM network into FM clusters of finite size and leads to the formation of a CSG state at temperatures T $\lesssim$ 8 K by promoting freezing of finite FM clusters in random orientations. Increasing strength of the single-ion magnetocrystalline anisotropy (and hence RMA) with decreasing temperature is taken to reflect an increase in the number of low-spin (LS) Co$^{3+}$ ions at the expense of that of high-spin (HS) Co$^{3+}$ ions. At intermediate temperatures (30 K $\lesssim T \lesssim$ 180 K), spin dynamics has contributions from the infinite FM network (fast relaxation governed by a single anisotropy energy barrier) and finite FM clusters (extremely slow stretched exponential relaxation due to hierarchical energy barriers).  
\end{abstract}

	
	

\maketitle
\section{Introduction}
Low-dimensional magnetic systems are susceptible to a variety of external perturbations, such as doping, magnetic field, and strain, etc.~\cite{Gon2019}. Their higher sensitivity to external stimuli compared to the 3D counterparts opens up room for new applications, $e.g.$, control of exchange bias in a single unit cell~\cite{Abd2020}. Moreover, in low-dimensional systems, control of spin moment can provide tremendous application opportunities for spintronics devices (layered cobaltates are potential candidates). LaCoO$_3$ (Co$^{3+}$), an often studied 3D compound, exhibits peculiar properties such as a metal-insulator transition, high-spin (HS) state to low-spin (LS) state transitions, magneto-electronic phase separation and ferromagnetic (FM) correlations~\cite{Jwu2003}. A fascinating cobaltate and its 2D analog, La$_2$CoO$_4$ (Co$^{2+}$) is an antiferromagnetic insulator (AFI)~\cite{Rao1988}, but not explored enough mainly due to its topotactic oxidation~\cite{Nem1998}. Its counterpart Sr$_2$CoO$_4$ (Co$^{4+}$) has a similar structure and shows ferromagnetism and half-metallicity~\cite{Pan2013}. In LaSrCoO$_4$ (an intermediate composition of the above-mentioned end compounds), La/SrCoO$_3$ 3D blocks are separated by rock-salt La/SrO layers.  

Depending upon how the energy-splitting $\Delta_{CF}$ between the $t_{2g}$ and $e_{g}$ orbitals, caused by the crystal-field, compares in magnitude with the exchange energy $J_{ex}$ associated with the Hund's rule coupling, the Co$^{3+}$ ion can exist in three different spin states~\cite{Abd2017}: the magnetically-active Co$^{3+}$ high-spin (HS; $t_{2g}^{4}e_{g}^{2}$, S = 2) state when $\Delta_{CF}$ $<$ $J_{ex}$, intermediate-spin (IS; $t_{2g}^{5}e_{g}^{1}$, S = 1) state when $\Delta_{CF}$ $\sim$ $J_{ex}$ and non-magnetic low-spin (LS; $t_{2g}^{6}e_{g}^{0}$, S = 0) state when $\Delta_{CF}$ $\gg$ $J_{ex}$. In the LaSrCoO$_4$ (LSCO) compound, the spin-state of Co$^{3+}$ ions has been a controversial issue. For instance, LSCO is reported~\cite{Wan2000,Hwu2010,Hol2008,Mer2010, Mer2011} to have a mixture of LS and (thermally-activated) HS states of the Co$^{3+}$ ions. Presence of homogeneous IS state of Co$^{3+}$ ions has also been claimed~\cite{Ang2008} and subsequently challenged~\cite{Sub2020}. In our previous study~\cite{Abd2017}, we probed the spin-states in La$_{2-x}$Sr$_x$CoO$_4$ (0.5 $\leq$ x $\leq$ 1) compounds and found existence of the HS and LS states together at room temperature, discarding the presence of IS state. First-principles calculations ~\cite{Hwu2010} support the view that mixed spin-states (HS and LS) of Co$^{3+}$ ions inhabit the ground state of LaSrCoO$_4$. 

Existence of mixed spin-states of the Co$^{3+}$ ions in LSCO is expected to have an important bearing on the magnetic and transport properties, since the super-exchange interaction between the HS Co$^{3+}$ ions mediated by the intervening LS Co$^{3+}$ ion can induce long-range FM order, while the superexchange interaction between the adjacent HS Co$^{3+}$ neighbors can give rise to antiferromagnetic (AFM) order. The competing FM and AFM exchange interactions, in turn, could result in a spin glass (SG) state. Conflicting reports ~\cite{Mor1997,Shi2006,Guo2016,Chi2006,Liu2005,And2006} about the nature of magnetism in LSCO, however, render the existing magnetic data inconclusive, as elucidated below. While a paramagnetic (PM)-SG transition at $T_{SG}$ $\sim$ 20 K \cite{Mor1997} ($T_{SG}$ $\sim$ 8 K \cite{Shi2006}) is construed as a transition from the HS to IS state, no PM-SG transition is observed down to 2 K (the temperature up to which an insulating PM state persists \cite{Chi2006}) and a broad hump in the thermomagnetic data, observed at $\simeq$ 250 K, is attributed to the ferromagnetic (La,Sr)CoO$_{3}$ impurity \cite{Chi2006}. This insulating PM ground state is taken to reflect \cite{Chi2006} the IS spin state of Co$^{3+}$ ions. In sharp contrast, two magnetic transitions, PM-FM transition at T$_{c}$ $\simeq$ 250 K \cite{Liu2005,And2006} and the SG-like transition at $T_{SG}$ $\sim$ 12 K \cite{Liu2005} or at $T_{SG}$ $\sim$ 60 K \cite{And2006}, have been reported. Considering that the spin and valence states of Co are inextricably linked in cobaltates, presence of valence states (Co$^{2+}$ and/or Co$^{4+}$) of Co, arising from the varying degree of oxygen off-stoichiometry in the LSCO samples used in such investigations, could be at the root of these apparent discrepancies.

In this work, the intrinsic magnetic behavior near the PM - FM and SG transitions is unraveled by ensuring close to a pure Co$^{3+}$ state in our LaSrCoO$_4$ sample. DC magnetization and frequency-dependent ac susceptibility (ACS) data, taken on the LaSrCoO$_4$ compound, unambiguously confirm the existence of two magnetic transitions: a thermodynamic (\textit{frequency-independent}) PM - FM phase transition at T$_{c}$ = 220.5 K and a \textit{frequency-dependent} cluster spin glass (SG) transition, approaching the value T$_{g}$ = 7.7 K in the zero-frequency limit. Both of these transitions are basically driven by random magnetic anisotropy (RMA). Increasing strength of the single-ion magnetocrystalline anisotropy (and hence RMA) with decreasing temperature is shown to be a manifestation of an increase in the number of low-spin (LS) Co$^{3+}$ ions at the expense of that of high-spin (HS) Co$^{3+}$ ions.

\section{Experimental method}
The details of sample synthesis are reported elsewhere~\cite{Abd2017}. Synchrotron X-ray diffraction (XRD) pattern was measured at room temperature using the BL-12 ADXRD beamline of RRCAT, India. To avoid possible fluorescence from Co, an x-ray photon energy of 7690 eV ($\lambda_{XRD}$ = 1.612 ~\AA), which lies below the absorption edge, was used for XRD measurements. X-ray absorption near edge spectra (XANES) were recorded in the fluorescence mode at the same beamline. Energy resolution at the Co K-edge energy was $\sim$ 0.7 eV. `Zero-field' (H = 0) and `in-field' (H = 80 kOe) electrical resistivities as functions of temperature were measured by four-probe method, using the Oxford cryostat. Magnetization measurements were carried out utilizing the 7 Tesla Quantum design MPMS 3 magnetometer. M(H) hysteresis loops at fixed temperatures were recorded following the `field-cooled' (FC) protocol. Magnetization was measured as a function of temperature, M(T), over the temperature range extending from 5 K to 300 K at fixed fields, employing the FC and `zero-field-cooled' (ZFC) protocols. The virgin M(H) curves were recorded after demagnetizing the sample and the superconducting magnet by setting field to zero in the oscillatory mode. For constructing the Arrott plots from the virgin M(H) isotherms, the applied field has been corrected for the demagnetizing field to arrive at the internal effective magnetic field~\cite{Kau1999} H$_{eff}$, in accordance with the relation, H$_{eff}$ = H$_{applied}$ - (4$\pi$N) M, where N is the demagnetizing factor, which was determined from the low-field portions of the M(H) isotherms taken at T $<$ T$_c$. ZFC relaxation measurement has been performed by cooling the sample down to 150 K in zero field and after waiting for 100 s (t$_w$), a static magnetic field of 50 Oe was applied and the time evolution of magnetization at 150 K was measured upto $\sim$ 8000 s. The real ($\chi'(\omega,T)$) and imaginary ($\chi''(\omega,T)$) components of ac magnetic susceptibility were measured in the temperature range of 2 - 300 K, at fixed frequencies (1.1 Hz - 941 Hz) of the ac driving field of rms amplitude, $h_{ac}$= 3 Oe employing the Quantum Design MPMS 3 system equipped with an ac susceptibility option.

\begin{figure}[hbt]
	\centering
	\includegraphics[width=0.68\textwidth]{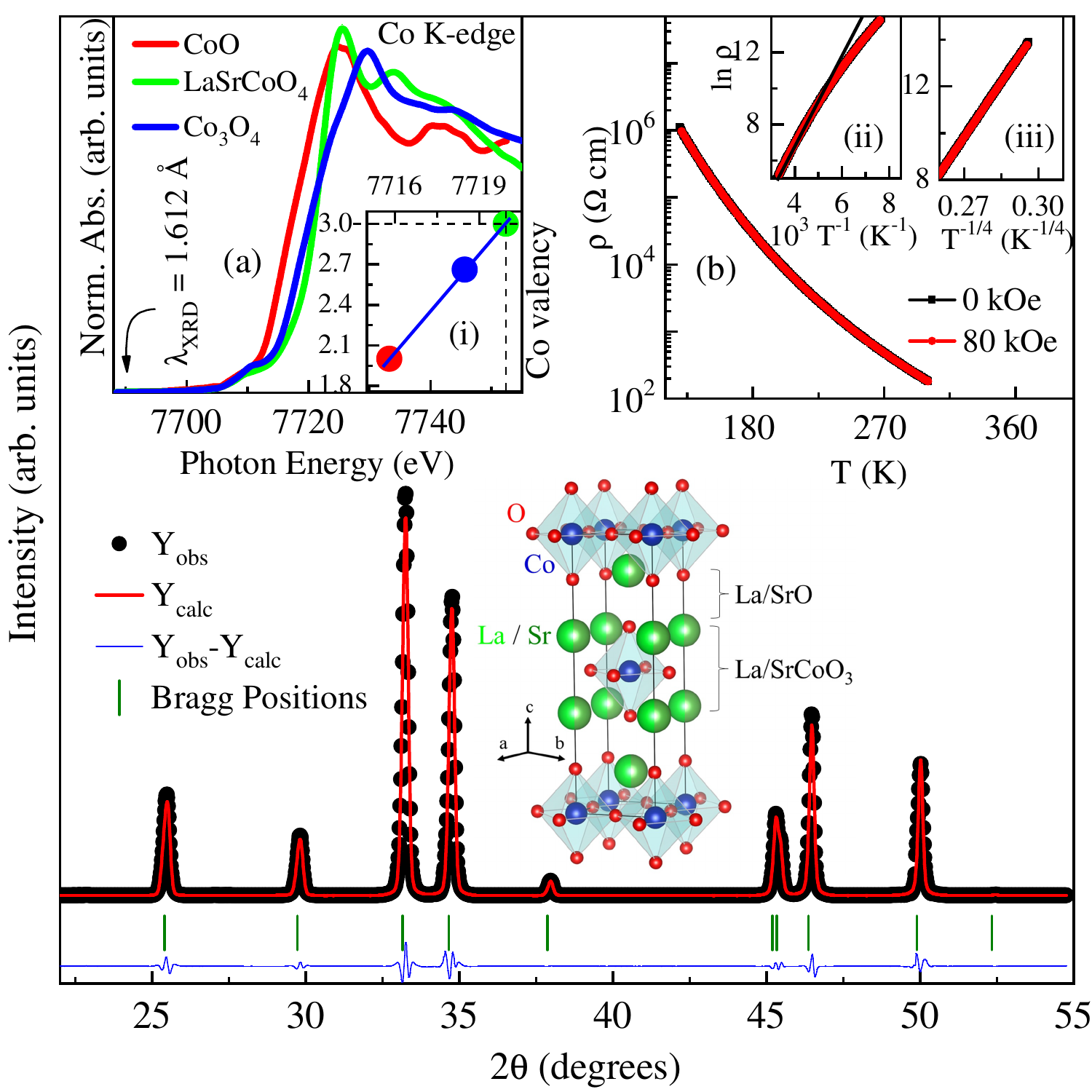}
	\caption{Room temperature synchrotron X-ray diffraction pattern of LaSrCoO$_4$ (LSCO) compound, measured at the x-ray photon energy of 7690 eV, along with the Rietveld fit. A Schematic of the unit cell shows that the structure consists of  blocks of La/SrCoO$_3$ separated by the La/SrO layers. Inset (a) compares the Co K-edge XANES of LaSrCoO$_4$ with those of the reference compounds CoO and Co$_3$O$_4$. Linear Co-valency versus Co K-edge photon energy plot in the inset (i) extrapolates to the Co-valency of +3 in the LSCO compound. Inset (b) displays the temperature variations of the `zero-field' ($\rho$(H = 0)) and `in-field' ($\rho$(H = 80 kOe)) resistivity. The insets (ii) and (iii) depict the $ln\rho$ versus $T^{-1}$ and $ln\rho$ versus $T^{-1/4}$ plots, respectively.}
	\label{1}
\end{figure}

\section{Results and discussion}

Fig.~\ref{1} shows room temperature synchrotron XRD pattern of the LaSrCoO$_4$ (LSCO) compound together with the Rietveld fit. Absence of any extra Bragg peaks, attributable to the impurity phases such as La$_2$CoO$_4$ (Co$^{2+}$) and Sr$_2$CoO$_4$ (Co$^{4+}$), confirms the single-phase nature of the LaSrCoO$_4$ sample. Rietveld refinement yielded the lattice parameters, corresponding to the tetragonal space group $I$4/$mmm$, as $a$ = 3.8025(3)~\AA~ and $c$ = 12.4985(2)~\AA~. Inset (a) compares the Co K-edge XANES data of LaSrCoO$_4$ with those of the reference compounds CoO and Co$_3$O$_4$. Linear Co-valency versus Co K-edge photon energy plot in the inset (i) yields the value +2.98(2) for the Co valence and thereby confirms the Co-valency of +3 in the LSCO sample. Data points, depicted by circles in the linear plot shown in inset (i), correspond to the first derivative of the absorption edge. 

Inset (b) of Fig.~\ref{1} clearly demonstrates semiconducting-like temperature variation of the `zero-field' ($\rho$(H = 0)) and `in-field' ($\rho$(H = 80 kOe)) resistivity in the LSCO sample. If, apart from Co$^{3+}$ ions, Co$^{2+}$ and Co$^{4+}$ ions were also present in the sample, double exchange, involving charge hopping, would have resulted in a metallic state (characterized by low resistivity, typically $\sim$ 10 - 100 $\mu\Omega~cm$, and $\rho$ increasing with temperature) as opposed to the observed semiconducting-like state with very large resistivity decreasing from $10^{6}~ \Omega~cm$ at 130 K to $10^{2}~ \Omega~cm$ at 310 K. Consistent with the conclusion drawn from the Co K-edge XANES data, this finding completely rules out the presence of Co$^{2+}$ and Co$^{4+}$ ions in our LSCO sample. Furthermore, negligibly small magnetoresistance (MR), [($\rho$(T,H = 80 kOe) - $\rho$(T,H = 0)]/$\rho$(T,H = 0), observed in the present case, sharply contrasts the reasonably large negative MR, expected in double-exchange ferromagnets. The insets (ii) and (iii) of Fig.~\ref{1}(b) bear out clearly that the processes such as the thermal activation of charge carriers across the band gap (described by expression $\rho$ $\sim$ $e^{E/k_{B}T}$) and Mott variable-range polaron hopping (represented by the expression $\rho$ $\sim$ $e^{(T_{0}/T)^{1/4}}$) \cite{Mott1968,Mott1993,Bit2020} operate within the temperature ranges 225 K - 300 K (T $>$ T$_{c}$) and 130 K - 220 K (T $\leq$ T$_{c}$), respectively. 

\begin{figure}[hbt]
	\centering
	\includegraphics[width=0.68\textwidth]{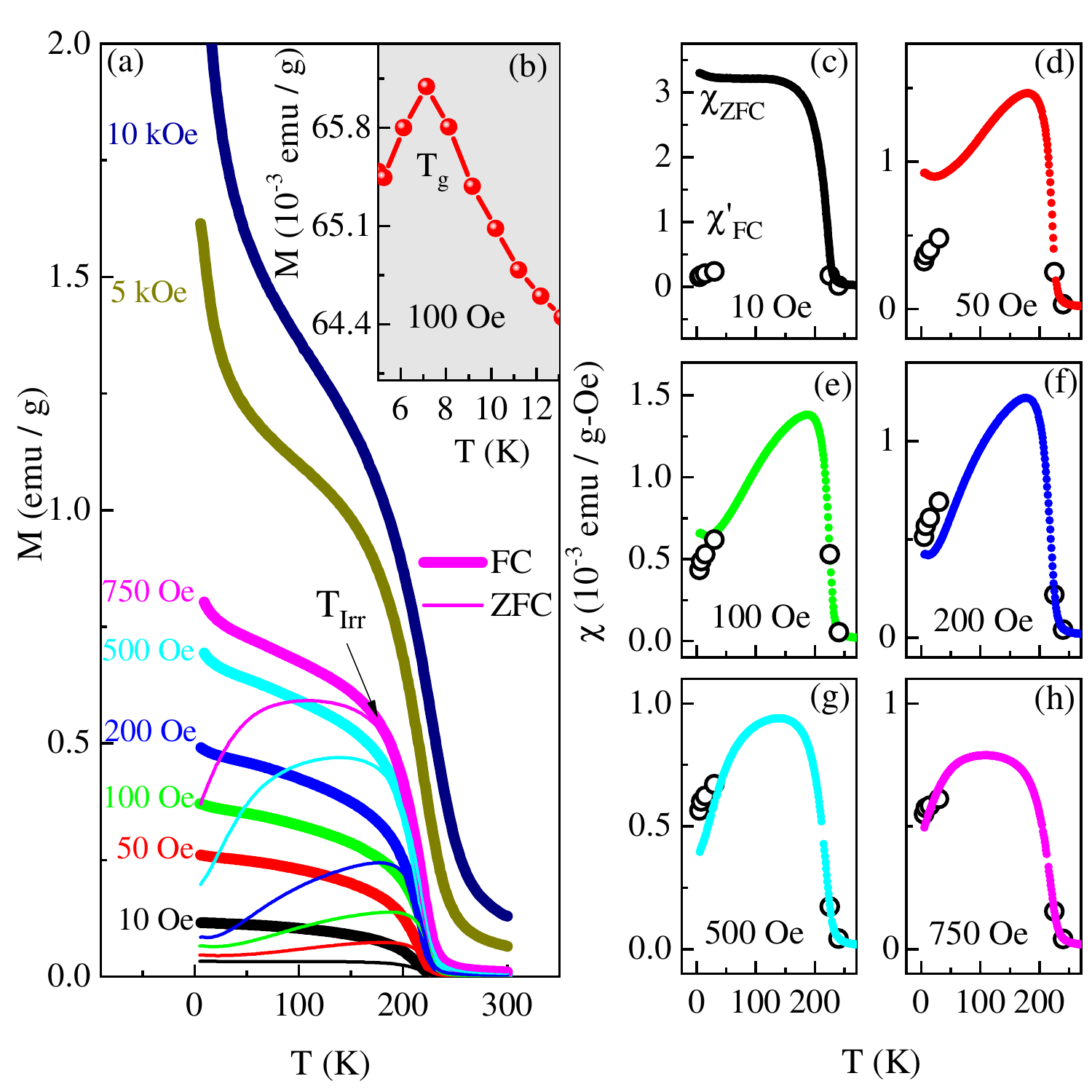}
	\caption{(a) Thermomagnetic M(T) curves taken at fixed magnetic fields. Inset (b) shows the enlarged view of ZFC M(T) at low temperatures which facilitates the observation of the spin glass transition at T$_g$. (c-h) Temperature variations of the susceptibilities measured (solid circles) and calculated (open circles) at the different fixed magnetic fields. }
	\label{2}
\end{figure}

After ensuring that the sample is of very good quality (free from the impurity phases), in the following, we attempt to unravel the \textit{intrinsic} magnetic behavior of the LaSrCoO$_4$ compound. Fig.~\ref{2}(a) displays the `field-cooled' (FC) and `zero-field-cooled' (ZFC) thermomagnetic curves, M(T), at fixed magnetic fields in the range 10 Oe - 10 kOe. A clear bifurcation between FC and ZFC M(T) is observed at low fields. Such bifurcations are indicative of either a spin-glass (SG) phase~\cite{Kri2004} or cluster spin-glass (CG) phase or magnetic anisotropy energy barriers or super-paramagnetism (SPM) ~\cite{Joy1998}. A closer look at the low-temperature ZFC M(T) data shows a clear cusp (inset (b) of Fig.~\ref{2}) at T$_g$ $\sim$ 8 K. Similar feature was observed previously~\cite{Guo2016,Liu2005,Shi2006} and denoted as a freezing temperature. An interesting point to note is that on cooling below T$_{Irr}$ (the temperature at which FC and ZFC M(T) bifurcate), M$_{FC}$ increases, which is taken to be a characteristic feature of cluster spin glass behavior in various systems~\cite{Fre2001}. The increase in M$_{FC}$ with decreasing temperature, even below T$_g$, points to the presence of a cluster spin glass (CSG) state~\cite{Kau1998} because M$_{FC}$ normally exhibits a plateau for canonical spin glass~\cite{Pej2000}. It is noteworthy to mention that, for a typical re-entrant SG systems, the irreversibility occurs far below T$_c$ while for CSG T$_{Irr}$ $\leq$ T$_c$~\cite{Muk1996}. In this connection, it is also reported~\cite{Jwu2003} that if the bifurcation temperature T$_{Irr}$ $>>$ T$_g$, the compound exhibits cluster spin glass type behavior while, for canonical SG~\cite{Myd2015}, the T$_{Irr}$ coincides with the T$_g$. Thus, we conclude that the signatures discussed above basically reflect the CSG state in LaSrCoO$_4$ at low temperatures. 

\begin{figure}[!]
	\centering
	\includegraphics[width=0.67\textwidth]{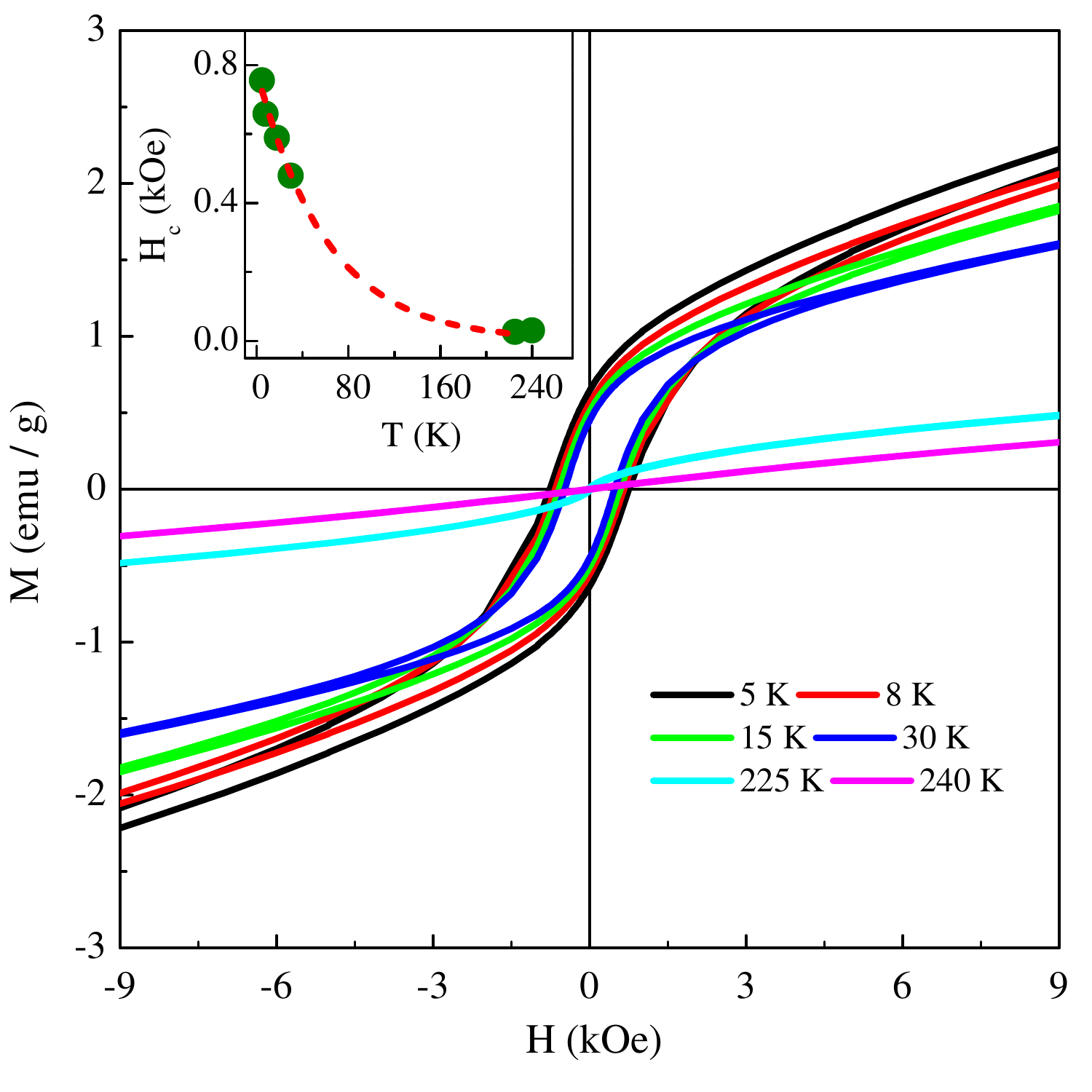}
	\caption{`Field-cooled' M(H) hysteresis loops. Inset shows the variation of H$_c$ with temperature.}
	\label{3}
\end{figure}

Next, we discuss the role of magnetic anisotropy as a possible source of bifurcation in M(T). Because of the prevalent elongated octahedra in the crystal structures similar to that of LaSrCoO$_4$, the pseudo orbital moment of Co$^{3+}$ HS, $\tilde{L}$ = 1~\cite{Hav2006}, prefers to lie in-plane and forces the spin moment also to lie within the plane~\cite{Csi2005} due to the spin-orbit coupling. Neutron diffraction studies on iso-structural compounds LaSrFeO$_4$~\cite{Qur2013} and half-doped La$_{1.5}$Sr$_{0.5}$CoO$_4$ ~\cite{Hel2009} have clearly borne out that, in these systems, magnetic moments are confined to the $ab$ plane but can rotate within the plane. This is indicative of an XY-type anisotropy and hence $c$-axis may not be the easy axis of magnetization. To ascertain whether or not magnetic anisotropy plays a role, we adopted the approach, proposed by Joy $et~al.$~\cite{Joy1998}, according to which, if the decrease in M$_{ZFC}$ at low temperatures is due to magnetocrystalline anisotropy, M$_{ZFC}$ should follow the relation, $\frac{M_{FC}}{H_{app}+H_c} \approx \frac{M_{ZFC}}{H_{app}}$. That this is indeed the case for the fields $H \geq$ 100 Oe, is evident from Fig.~\ref{2}(e - h). However, at $H \leq $ 50 Oe, $\chi_{ZFC}$ ($=\frac{M_{ZFC}}{H_{app}}$) strongly deviates from $\chi^{'}_{FC}$ ($=\frac{M_{FC}}{H_{app}+H_c}$) (see Fig.~\ref{2}(c, d)) in the low temperature region. The demagnetization-limited behavior of $\chi_{ZFC}$ at $T < T_{c}$ for $H =$ 10 Oe, suggests that the shape anisotropy also becomes important at low fields $H \leq $ 50 Oe.  Note that, in the calculation of $\chi^{'}_{FC}$, we have used the coercive field, H$_c$(T), data shown in the inset of Fig.~\ref{3}.  

Fig.~\ref{3} presents the M(H) hysteresis loops at fixed temperatures, taken in the `field-cooled' (H = 70 kOe) mode at temperatures below and above the PM-FM transition temperature, T$_c$ = 220.5 K, (corresponding to the dip in dM$_{ZFC}$(T)/dT at low fields H = 10 - 100 Oe, not shown here). With temperature decreasing from T $\gg$ T$_c$, where the sample is in the PM state and as such the M(H) isotherms are linear, the coercive field, H$_c$, increases from zero at T = T$_c$ = 220.5 K to about 800 Oe at 5 K; the rate of increase in H$_c$ picks up for temperatures below 30 K, as is evident from the inset of Fig.~\ref{3}. A steep increase in H$_c$ as the temperature falls below 30 K is indicative of an increase in the random magnetic anisotropy (RMA). The observation that magnetization does not saturate even in fields as high as 70 kOe at low temperatures (Fig.5(a)) and the presence of RMA (as inferred from the critical-point analysis in the following section) support the existence of a CSG state at low temperatures. 

\begin{figure}[!]
	\centering
	\includegraphics[width=0.66\textwidth]{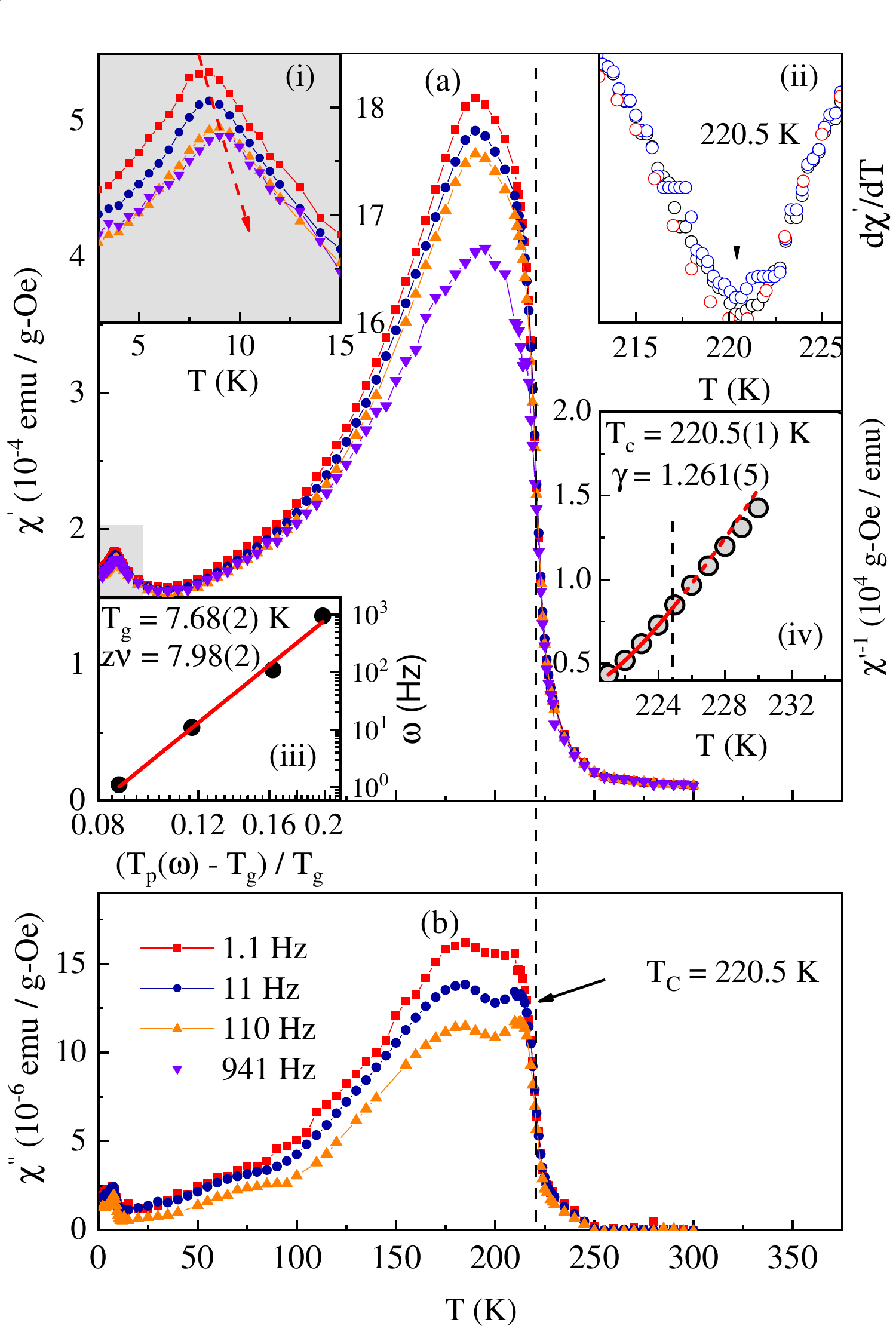}
	\caption{Temperature dependence of (a) `in-phase' component, $\chi'$, and (b) `out-of-phase' component, $\chi''$ of $\chi_{ac}$ at the different frequencies (1.1 to 941 Hz) measured at an ac driving field of rms amplitude 3 Oe. Inset (i) of panel (a) shows a zoomed view of the frequency dependence of the peak in $\chi'$(T) at around $\sim$ 8 K while the inset (ii) depicts the d$\chi^{'}$/dT $vs.$ T at different frequencies. The linear fit through the $\omega$ versus $({T_{p}(\omega)- T_g}) / T_g$ data (solid circles) in the inset (iii) testifies to the validity of the critical slowing down model. Inset (iv) displays the fit (red curve) to the inverse ac susceptibility, $\chi'^{-1}$(T), data (open circles) above $T_c$ = 220.5 K in the range 221 K - 225 K, based on Eq.(2). This fit, when extended to 230 K (dashed curve), demonstrates that the $\chi'^{-1}$(T) data start deviating from the fit for T $>$ 225 K. }
	\label{4}
\end{figure}

The panels (a) and (b) of Fig.~\ref{4} display the temperature variations of the real ($\chi'$) and imaginary ($\chi''$) parts of ac magnetic susceptibility, respectively. While the inset (i) of panel (a) highlights the frequency-induced shift in the peak in $\chi'$(T) at T = $T_{p}$ $\simeq$ 8 K, the inset (ii) clearly brings out the frequency-independence of the dip in the d$\chi'$/dT $vs.$ T plots at $T_c$ = 220.5 K, which corresponds to the inflection point in $\chi'$(T). The frequency-independent value of $T_c$ = 220.5 K confirms the true thermodynamic nature of the FM - PM phase transition at $T_c$.

The shift in $T_p$ per decade of frequency ($\Delta T_p / \{T_p \Delta(log_{10} \omega)  \}$ = 3.2 $\times$ 10$^{-2}$) is an order of magnitude greater than that ($\simeq$ 10$^{-3}$)) observed \cite{Myd2015} in canonical spin glasses but is typical of the cluster spin glasses \cite{Myd2015,Bit2012}. The frequency dependence of the peak temperature, $T_{p}(\omega)$, is analyzed in terms of the following expression given by the `critical slowing down' model \cite{Myd2015,Bit2012}

\begin{equation}
\centering
\omega=\omega_0 \left[\frac{T_{p}(\omega)-T_g}{T_g}\right]^{z\nu}
\label{critical}
\end{equation}

where $\tau_o = 2\pi/\omega_0$ is the relaxation time due to the correlated spin dynamics, $T_g$ is the spin-glass transition temperature in the zero-frequency limit, $\nu$ is the spin-spin correlation length critical exponent and $z$ is the dynamical critical exponent. The linear fit through the $\omega$ versus $({T_{p}(\omega)- T_g}) / T_g$ data (solid circles) in the inset (iii) of Fig.~\ref{4}, based on the Eq.~(\ref{critical}) with the parameter values $\tau_o $ = 2.49(1) $\times$ 10$^{-8}$ s, $T_g$ = 7.68(2) K and the product $z\nu$ = 7.98(2), testifies to the validity of the critical slowing down model. Note that, in the canonical spin glasses, relaxation time, $\tau_o$, is typically of the order of 10$^{-12}$ s \cite{Myd2015,Bit2012}. Several orders of magnitude larger $\tau_o$ (i.e., much slower spin dynamics), in the present case, further confirms the existence of a cluster spin glass ground state.

Next, we focus our attention on the FM - PM transition at $T_c$ = 220.5 K. In the critical region above $T_c$, the inverse of measured ac susceptibility, $\chi_{ac}^{-1}$(T) $\equiv \chi'^{-1}$(T), is related to the inverse intrinsic susceptibility, $\chi_{int}^{-1}$(T) as $\chi_{ac}^{-1}$(T) = 4$\pi$ N + $\chi_{int}^{-1}$(T), where N is the demagnetizing factor. As the temperature approaches $T_c$ from above, $\chi_{int}$ diverges (or equivalently, $\chi_{int}^{-1}$ goes to zero) at $T_c$ in accordance with the relation

\begin{equation}
\centering
\chi_{ac}^{-1}(T) = 4 \pi N + A \left[\frac{T - T_c}{T_c}\right]^{\gamma}
\label{critical ACS}
\end{equation}   

where $A$ and $\gamma$ are the critical amplitude and critical exponent, respectively. Inset (iv) of Fig.~\ref{4} depicts the best theoretical fit (continuous curve) to the $\chi_{ac}^{-1}(T)$ data, based on the Eq.~(\ref{critical ACS}), yields $T_c$ = 220.5 K, $A$ = 5.9(1) $\times$ 10$^5$ and $\gamma$ = 1.261(5). This value of $\gamma$ will be put into a proper context later in the text.

\begin{figure}[!]
	\centering
	\includegraphics[width=0.42\textwidth]{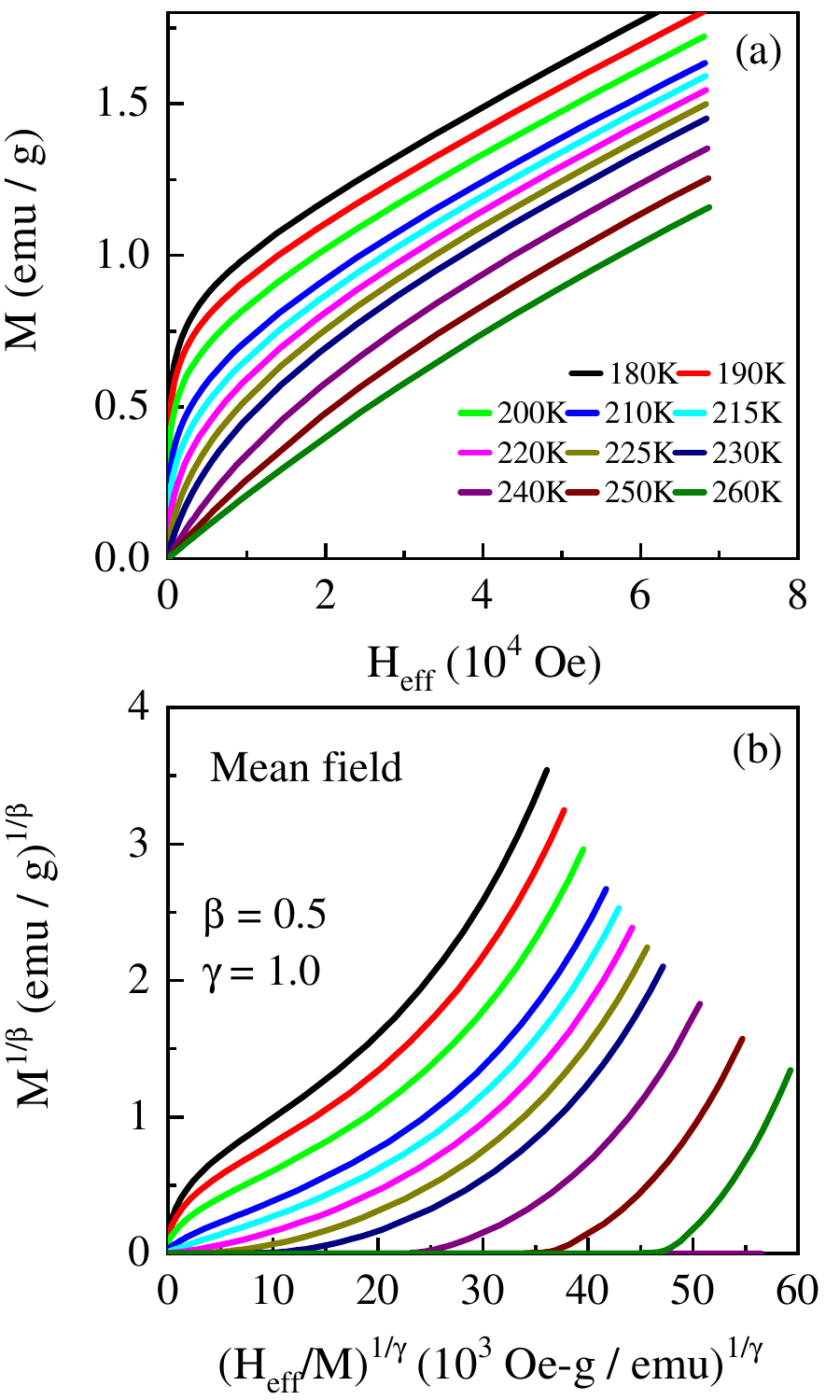}
	\caption{(a) M(H) isotherms at and around $T_{c}$. (b) The modified Arrott ($ M^{1/\beta}$ $vs.$ $(H/M)^{1/\gamma}$) plots with the choice of the critical exponents $\beta$ and $\gamma$ given by the mean-field theory.}
	\label{5}
\end{figure}

\begin{figure}[b]
	\centering
	\includegraphics[width=0.47\textwidth]{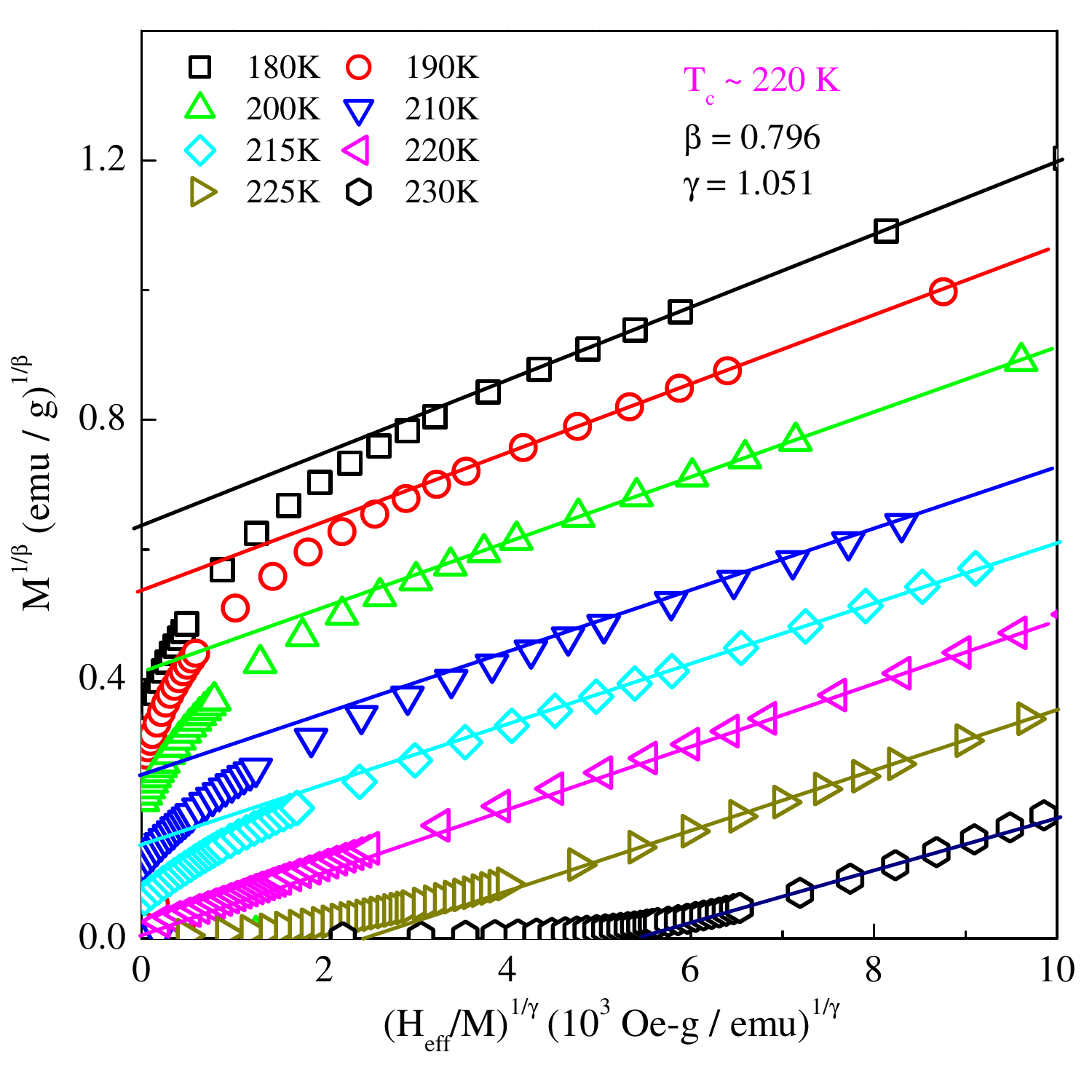}
	\caption{Arrott plots constructed using the critical exponent values $\beta$ = 0.796 and $\gamma$ = 1.051.}
	\label{6}
\end{figure}

It is well-known that the critical-point (CP) analysis of magnetization, M(T,H), data taken in the critical region near T$_c$ provides a powerful means to unravel the true nature of magnetic ordering prevalent in a spin system. As elucidated below, the CP analysis makes use of the critical exponents $\beta$, $\gamma$ and $\delta$ for the spontaneous magnetization (order parameter), `zero-field' susceptibility and the critical M(H) isotherm, (characterizing a continuous FM-PM phase transition at T$_c$), that are defined as ~\cite{Kau1985}

\begin{equation}
\centering
M_s = B~|\epsilon|^{\beta} \qquad T < T_c, ~~~~~~~~ H \rightarrow 0
\label{ms}
\end{equation}

\begin{equation}
\centering
\chi_0 = A ~\epsilon^{-\gamma}\qquad T > T_c, ~~~~~~~~H \rightarrow 0
\label{chi}
\end{equation}

\begin{equation}
\centering
M = M_0 ~~ H^{1/\delta} \qquad T = T_c
\label{del}
\end{equation}

where $\epsilon = (T - T_c)/T_c$, A, B and M$_0$ are the critical amplitudes~\cite{Kau1985,Kha2010}. Diverse systems having exactly the same values for the critical exponents $\beta$, $\gamma$ and $\delta$, fall into a single universality class. Universality class, in turn, is governed by the order parameter (spin) dimensionality ($n$) and spatial dimensionality ($d$)~\cite{Kau1985, Rei1995} as long as the interaction coupling the spins is of short range. The presence of different valence and spin states in the present system is expected to have a profound effect on the critical behavior. This expectation called for a detailed study of the critical behavior of LaSrCoO$_4$ at temperatures in the vicinity of the FM-PM phase transition. To this end, we have analyzed the virgin M(H) isotherms in the critical region (shown in Fig.~\ref{5}(a)) using the scaling equation of state (SES) method, detailed in the reference ~\cite{Kau1985}. Instead of following the customary practice of using, at first, the Arrott~\cite{Arr1957} SES, we use the generalized form of the Arrott SES, given by Arrott and Noakes~\cite{Arr1967} (AN), i.e., 

\begin{equation}
\centering
(H/M)^{1/\gamma} = a [(T - T_c)/T_{c}] + b M^{1/\beta}
\label{5e}
\end{equation}

\begin{figure*}[!]
	\centering
	\includegraphics[width=0.79\textwidth]{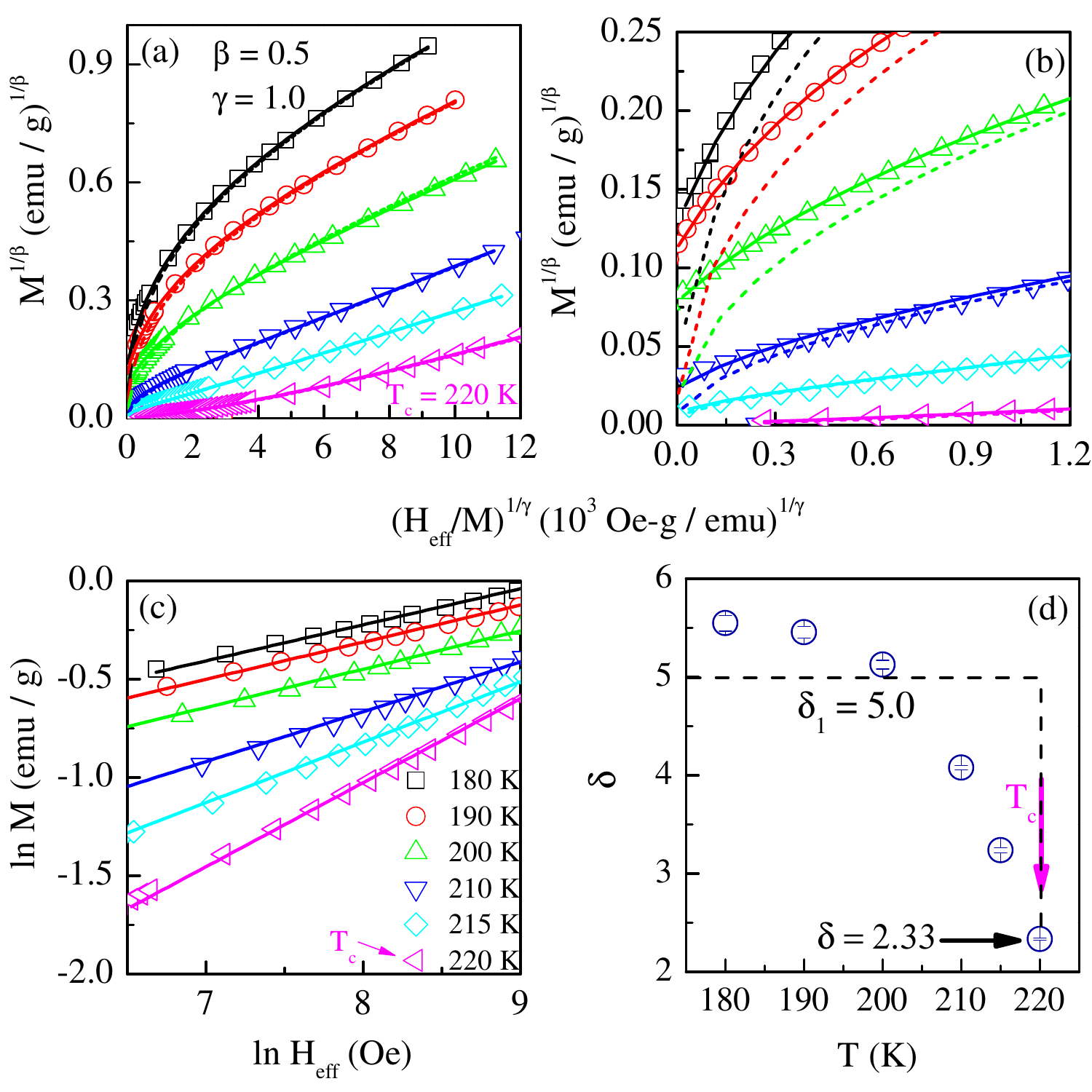}
	\caption{(a) Theoretical fits to the Arrott plot ($i.e.$, $M^{1/\beta}$ versus $(H/M)^{1/\gamma}$ plot with mean-field exponents $\beta = 0.5$ and $\gamma = 1$) isotherms, based on the scaling equation of state, Eq.~(\ref{ahpet}), predicted by the random anisotropy model. (b) Enlarged view of the data and fits at fields nearing zero. (c) The linear log$M$ versus log$H_{eff}$ plots at T$\leq$ T$_c$. (d) Critical exponent $\delta_{1}$ at different temperatures below T$_c$ and the critical exponent $\delta$ at T$_c$, obtained from the linear fits to the log$M$ versus log$H_{eff}$ data shown in (c).}
	\label{7}
\end{figure*}

to arrive at the correct choice of the critical exponents that makes the AN, $M^{1/\beta}$ versus $(H/M)^{1/\gamma}$, plots a set of parallel straight lines in the critical region near T$_c$. The temperature at which the linear AN plot isotherm passes through the origin marks the T$_c$. To begin with, we ascertain if the values of critical exponents $\beta$ and $\gamma$, theoretically predicted for any universality class, make the AN plot isotherms at temperatures close to T$_c$ linear. For example, in Fig.~\ref{5} (b) it is evident that the mean-field (MF) critical exponent values $\beta$ = 0.5 and $\gamma$ = 1 do not yield linear AN isotherm in the critical region. The data presented in Fig. S1 of the supplementary material demonstrate that, like the MF critical exponent values, none of the universality class critical exponent choices: $\beta$ = 0.365, $\gamma$ = 1.386 for three-dimensional (3D) Heisenberg; $\beta$ = 0.345, $\gamma$ = 1.316 for 3D-XY; $\beta$ = 0.325, $\gamma$ = 1.241 for 3D Ising and $\beta$ = 0.25, $\gamma$ = 1.0 for mean-field tricritical, results in the linear AN plot isotherms.

 Furthermore, regardless of the choice of the critical exponents $\beta$ and $\gamma$, extrapolation of the high-field portions of the AN isotherms to $H = 0$ does not give any intercept on the $M^{1/\beta}$ axis. No intercept on the ordinate axis implies no spontaneous magnetization and hence absence of long-range FM order. This result prompted us to look for even the non-universal exponent values that could lead to the desired linear AN plot isotherms.     

By varying the values of $\beta$ and $\gamma$ in Eq.~(\ref{5e}), we succeeded in making the AN plot isotherms nearly straight over the field range 1.5 kOe $\lesssim H \lesssim$ 10 kOe with the critical exponent values, $\beta$ = 0.796, $\gamma$ = 1.051. Note that the field range for the linear AN isotherms is much wider in the immediate vicinity of T$_c$. From the modified Arrott plot (Fig.~\ref{6}), it is evident that the extrapolation of the linear portions of the AN isotherms to H = 0 yields finite spontaneous magnetization, $M_{s}$, (inverse susceptibility, $\chi^{-1}$) below (above) T$_c$ and both $M_{s}$ and $\chi^{-1}$ are zero at  the transition temperature T$_c$ $\simeq$ 220 K. This value of T$_c$ is very close to that deduced from the ac susceptibility data. Using the critical exponent values, $\beta$ = 0.796 and $\gamma$ = 1.051, in the Widom scaling relation~\cite{Lag2012} $\delta = 1 + \gamma/\beta$, we obtained the critical isotherm critical exponent $\delta$ as 2.32. To crosscheck this value, the relation (Eq.~(\ref{del})) is used to fit the virgin M(H) isotherm at T$_c$ $\simeq$ 220 K. Fig.~\ref{7}(c) shows the critical M(H) isotherm plotted as the ln M versus ln H$_{eff}$ plot. As expected from Eq.~(\ref{del}), this log-log plot is linear with inverse slope $\delta$ = 2.35(5). This value of $\delta$ is in excellent agreement with that obtained using the Widom scaling relation. 

\begin{figure}[t]
	\centering
	\includegraphics[width=0.4\textwidth]{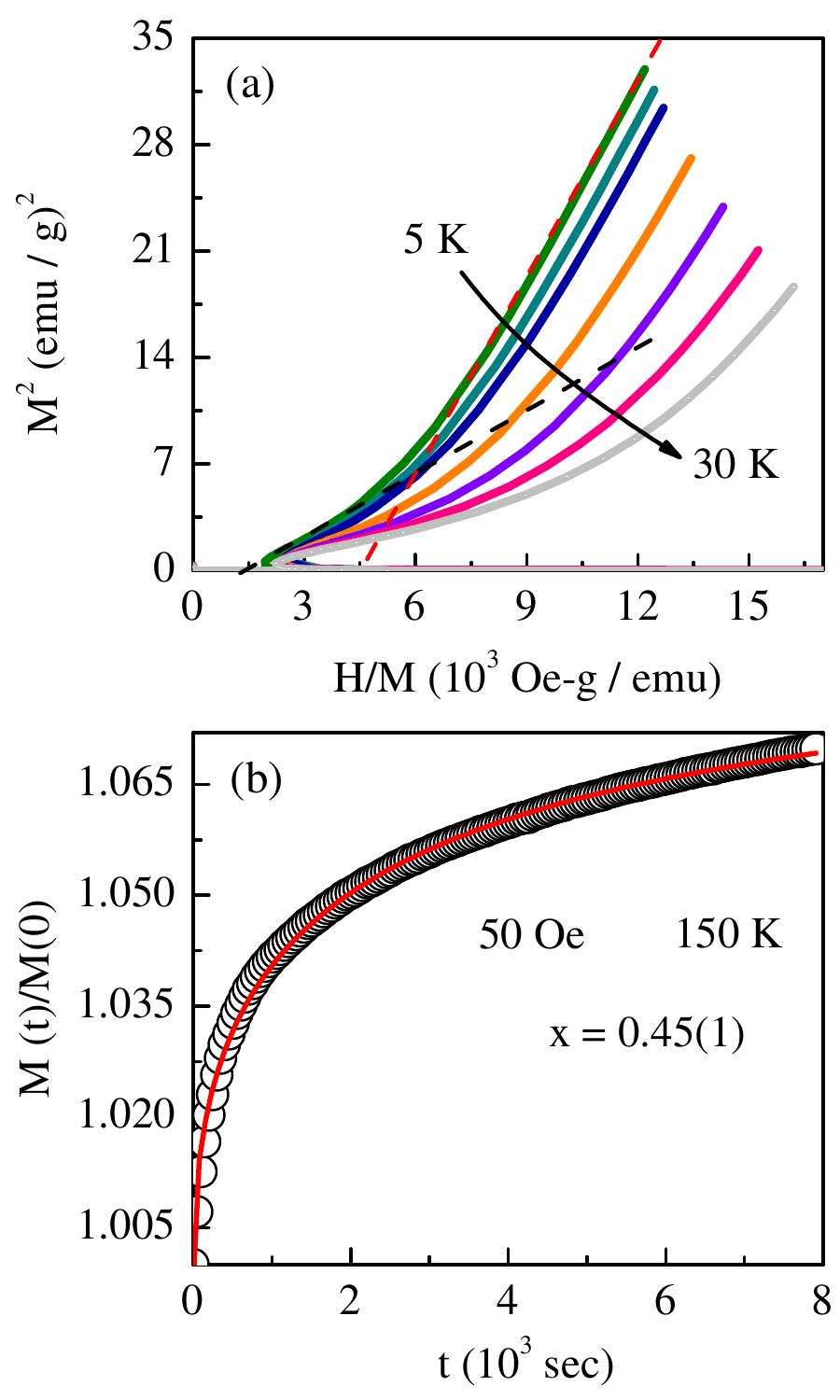}
	\caption{(a) M$^{2}$ versus (H/M) plots in the low (5-30 K) temperature region demonstrate that no spontaneous magnetization, M$_s$, exists at such temperatures. (b) Time evolution of the ZFC magnetization, M(t), at 150 K measured in a static magnetic field of 50 Oe switched on with a time delay of 100 s. Note that M(t) is normalized to its value at the time when the field was applied. The solid red curve through the data points (open circles) represents the fit to the stretched exponential function.}
	\label{8}
\end{figure}

Despite a close agreement between the value of $\delta$ directly determined from the critical isotherm and that from the exponent values $\beta$ = 0.796 and $\gamma$ = 1.051 via the Widom scaling relation, two basic questions arise: (i) how does one reconcile with the non-universal critical exponent values, and (ii) what causes the deviations from the linear AN isotherms at sufficiently low fields. Considering that the random magnetic anisotropy (RMA) model \cite{Aha1980} yields nonlinear mean-field AN isotherms at low fields, it is worth investigating if the SES form given by Aharony and Pytte~\cite{Aha1980} helps resolve the above issues. Note that nonlinear mean-field isotherms have been observed in the past in a number of amorphous rare earth - transition metal systems and explained ~\cite{Bar1985} in terms of the Aharony and Pytte model. 

Neglecting the critical fluctuations, Aharony and Pytte (AP) make the following predictions for a ($d$, $n$) spin system with RMA. (I) The M(H) isotherms follow the relation $H \sim M^{1/\delta}$ at $T_c$ and $H \sim M^{1/\delta_1}$ at all temperatures below $T_c$, with the exponents $\delta$ and $\delta_1$ given by
\begin{equation}
\centering
\delta = \frac{10-d}{6-d}, ~~~\delta_1 = \frac{8-d}{4-d}
\label{expo}
\end{equation}

For a three-dimensional ($d$ = 3) system, Eq.~(\ref{expo}) gives $\delta$ = 7/3 = 2.33 and $\delta_1$ = 5.0. While the theoretical estimate $\delta$ = 2.33 is in perfect agreement with the presently determined value $\delta$ = 2.35(5), the observed $\delta_1$ values do not reproduce the theoretically-predicted abrupt jump in $\delta_1$ at T $\simeq$ $T_c$ to the value $\delta_1$ = 5.0, which remains constant for T $<$ $T_c$ (as evidenced in Fig.~\ref{7}(d)). This discrepancy between theory and experiment can be traced back to a total neglect of critical fluctuations of the order parameter, leading to Eq.~(\ref{expo}). (II) According to the AP model, the scaling equation of state has the following form \cite{Aha1980} for a $d$ = 3, $n$ = 3 system with random uniaxial and cubic anisotropies  


\begin{equation}
\centering
M^{1/\beta} = \left.
\left[\left(\frac{H}{M}\right)^{1/\gamma} - a\frac{(T - T_c)}{T_c} \right] \middle /b \left[1 + c \left(\frac{H}{M}+ 4~\kappa~M^{2}\right)^{- 1 / 2}\right]
\right.
\label{ahpet}
\end{equation}


where $\beta$ = 0.5, $\gamma$ = 1.0, $c \sim (D/J)^2$, $D$ and $\kappa$ are a measure of random uniaxial and cubic anisotropies, respectively, and $J$ signifies the exchange interaction. Eq.~(\ref{ahpet}) permits a clear distinction between the cases: (i) $\kappa$ = 0 when only the random uniaxial anisotropy exists, and (ii) $\kappa$ $\neq$ 0 when, in addition to the random uniaxial anisotropy, cubic anisotropy is present. The low-field $M_{T}(H)$ data, plotted in the form of the Arrott plot (i.e., $M^{1/\beta}$ $vs.$ $(H/M)^{1/\gamma}$ plot with mean-field exponents $\beta = 0.5$ and $\gamma = 1$) isotherms, and the theoretical fits based on the AP SES, Eq.~(\ref{ahpet}), for the cases $\kappa$ = 0 (dashed curves) and $\kappa$ $\neq$ 0 (continuous curves), are displayed in Fig.~\ref{7} (a). Fig.~\ref{7} (b) provides an enlarged view of the data and fits at fields not very far from zero. Evidently, the AP SES correctly reproduces the observed change in the curvature of the Arrott plot isotherms from concave-downward to concave-upward at T$_c$. Furthermore, superiority of the $\kappa$ $\neq$ 0 fits over those corresponding to $\kappa$ = 0, particularly at fields close to zero, asserts that the inclusion of cubic anisotropy is necessary. Consequently, in conformity with the observed behavior, an extrapolation to $H$ = 0 gives finite intercepts on the ordinate (finite M$_{s}$) for T $<$ T$_{c}$ and on the abscissa (finite $\chi^{-1}$) for T $>$ T$_{c}$; both M$_{s}$ and $\chi^{-1}$ go to zero at T = T$_{c}$. However, due to sizable gradients of the AP isotherms at very low fields ($H \rightarrow$ 0) for temperatures on either side of T$_{c}$, the intercepts on the ordinate and abscissa depend on the range of $H$ chosen for extrapolation. Thus, the M$_{s}$(T) and $\chi^{-1}$(T), so obtained, are not accurate enough for the determination of critical exponents $\beta$ and $\gamma$. In sharp contrast, the `zero-field' ac susceptibility (ACS) does not suffer from such extrapolation errors. The value $\gamma$ = 1.261(5) of the susceptibility critical exponent, determined from the ACS data, is thus reliable. The deviation of this value of $\gamma$ from the mean-field value, $\gamma$ = 1.0, underscores the importance of critical fluctuations (neglected in the AP RMA model). Thus, the RMA model, due to Aharony and Pytte~\cite{Aha1980}, captures the main physical essence in that the curvature in the mean-field Arrott isotherms at low fields has its origin in the random uniaxial and cubic anisotropies, but fails to yield correct values for the critical exponents $\beta$ and $\gamma$.

The evidence presented so far asserts that (i) at temperatures in range 180 K $\leq$ T $\leq$ T$_c$ = 220.5 K, RMA is too weak to destroy long-range FM order, and (ii) the transition to a cluster spin glass (CSG) state occurs at T$_g$ = 7.7 K. It remains to be verified if, in the intermediate temperature range, a mixed state is formed in which long-range FM order coexists with the CSG order. In this connection, Fig.~\ref{8}(a) makes it obvious that, in the low temperature region (T $\lesssim$ 30 K), the Arrott (M$^{2}$ versus H/M) plots, when extrapolated to H = 0, do not yield any intercept on the ordinate axis regardless of the field-range chosen for extrapolation. It immediately follows that no spontaneous magnetization, and hence no long-range FM order, exists at these temperatures. Concurrently, in the same temperature range T $\lesssim$ 30 K, a steep rise in RMA (reflected in a sharp increase in the coercive field, as noticed in the inset of Fig.~\ref{3}) occurs. Within the framework of the RMA models \cite{Aha1980,Chud1986}, sizable magnitude of RMA at T $\lesssim$ 30 K can account for both the breakdown of long-range FM order \cite{Aha1980} and the freezing of spin clusters in random orientations at T = T$_g$, leading to the formation of cluster (correlated) spin glass state (with a finite FM correlation length \cite{Chud1986}) at temperatures below T$_g$.

With a view to gain more insight into the nature of magnetism at temperatures intermediate between the spin glass transition temperature T$_g$ $\simeq$ 8 K and T$_c$ $\simeq$ 220.5 K, the time (t) evolution of the ZFC magnetization, $M_{ZFC}$(t), at 150 K was measured in a static magnetic field of 50 Oe switched on after a fixed waiting time, $t_{w}$, ranging between $10^{2}$ and $10^{4}$ s. The $M_{ZFC}$(t) data, so obtained, were fitted to the stretched exponential function given by expression

\begin{equation}
\centering
M(t) = M_0 - M_r~exp(-(t/t_r)^x)
\label{relax}
\end{equation}

where $M_0$ represents the FM component while $M_r$ and $t_r$ are the spin-glass magnetization component and the characteristic SG relaxation time, respectively~\cite{Abd2019}. The value of exponent $x$ reflects the type of energy barriers present. For instance, $x$ = 1 describes the relaxation behavior of a FM system with a single anisotropy energy barrier separating the ZFC and FC states ~\cite{Kau2012}. $x$ $\simeq$ 0.5, on the other hand, is associated with the stretched exponential relaxation, caused by hierarchical energy barriers in spin glasses~\cite{Myd2015}. From the fit (solid red curve through the data points), based on the above expression (Eq.~(\ref{relax})), to the $M(t)/M(0)$ data (Fig.~\ref{8}(b)), we obtain $x = 0.45(1)$, $M_0/M(0)$ = 1.083(1), $M_r/M(0)$ = 0.086(1) and $t_r$ = 2252 sec, where M(0) = 1.0348 is magnetization value, recorded at the time $t_{w}$ when the field was applied. The determined value of $x$ is in good agreement with those generally reported for spin glass systems~\cite{Myd2015}. This result provides an evidence for the coexistence of a glassy (a cluster spin glass-like) phase with ferromagnetic order at intermediate temperatures. Note that, within the uncertainty limits, the above-mentioned parameter values seem to be insensitive to the choice of $t_{w}$ presumably because $M_r$ constitutes a tiny fraction of the total magnetization at the measurement temperature of 150 K, which lies well above T$_g$ $\simeq$ 8 K. 

\begin{figure}[t]
	\centering
	\includegraphics[width=0.8\textwidth]{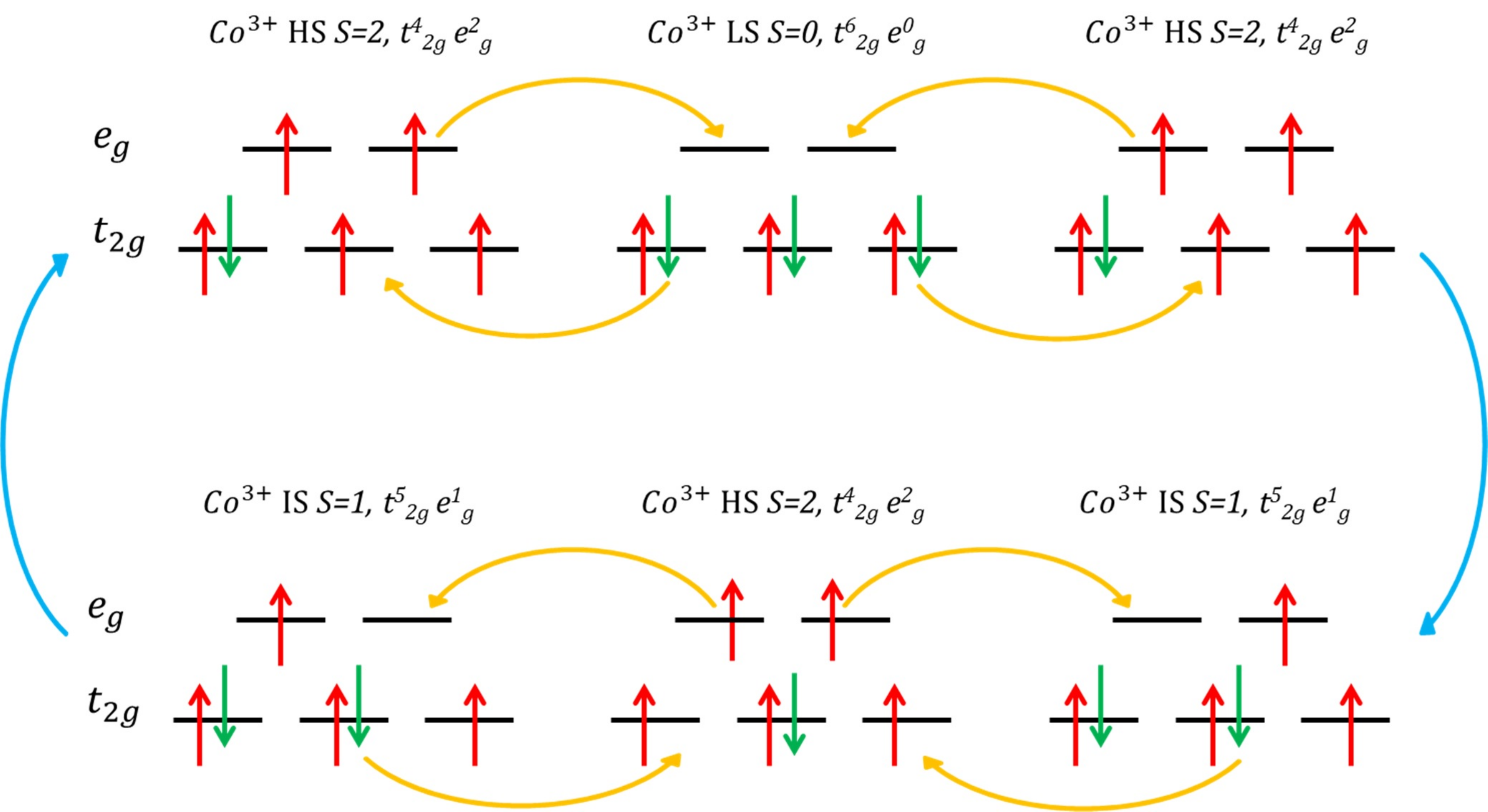}
	\caption{Schematic showing the two interchangeable spin state configurations involving correlated hopping of $t_{2g}$ and $e_g$ electrons between the HS and LS states in the ground state and HS and IS states in the excited state.}
	\label{9}
\end{figure}

Finally, an attempt is made to ascertain how the present results can be understood in terms of the HS/IS/LS spin states of Co$^{3+}$ ions in the LaSrCoO$_{4}$ compound. At first, we recall that the single-ion magneto-crystalline anisotropy (MCA) has its origin in the combined action of the crystal-field interaction (CFI) and spin-orbit coupling (SOC). Since the LS state is associated with S = 0, the LS Co$^{3+}$ ions, neither in isolation nor as neighbors, contribute to spin magnetic moment and MCA (due to the lack of SOC). However, in the case of LS Co$^{3+}$ ions with HS neighbors, the spin-state configurations, fluctuating ~\cite{Mer2011, Mer2010} between the Co$^{3+}$ HS - O$^{2-}$ - Co$^{3+}$ LS - O$^{2-}$ - Co$^{3+}$ HS and Co$^{3+}$ IS - O$^{2-}$ - Co$^{3+}$ HS - O$^{2-}$ - Co$^{3+}$ IS configurations (refer to the schematic diagram shown in Fig.~\ref{9}), can lead to FM superexchange interaction between the spins of Co$^{3+}$ HS ions. This process involves correlated hopping~\cite{Mer2011,Mer2010} of $t_{2g}$ and $e_{g}$ electrons between Co$^{3+}$ LS and its Co$^{3+}$ HS neighbors without any charge transfer as shown in Fig.~\ref{9} and results in a finite effective S at the LS Co$^{3+}$ ion site. Finite S, in turn, enables SOC and ensures a significant contribution to spin magnetic moment. In conformity with the observed steep rise in MCA as the temperature drops below $\sim$ 80 K, MCA increases with decreasing temperature because LS Co$^{3+}$ ions with SOC and much stronger CFI (and hence MCA) increase ~\cite{Mer2010, Mer2011, Seh1995} in number at the expense of HS Co$^{3+}$ ions. In sharp contrast, arguments based on the mixed HS-IS or IS ground state (similar to those presented above), at best, lead to a much slower increase in MCA with lowering temperature because of their low CFI relative to LS. A competition between the above-mentioned FM interaction and the AFM interaction between the spins of neighboring Co$^{3+}$ HS ions, caused by the Co$^{3+}$ HS - O$^{2-}$ - Co$^{3+}$ HS superexchange, give rise to a spin glass (SG) ground state.

\section{Summary and Conclusion}

In the LaSrCoO$_4$ compound, as functions of temperature, `zero-field-cooled' (ZFC) dc susceptibility, $\chi_{dc} = M/H$, at low fields (H $\leq$ 100 Oe), and ac susceptibility, $\chi_{ac}$, (ACS), at different frequencies of the driving ac field of rms amplitude $h_{ac}$ = 3 Oe, exhibit peaks at around 200 K and 8 K. While the low-temperature peak in $\chi_{ac}$(T) shifts to higher temperatures with increasing frequency, the inflection-point in $\chi_{ac}$(T) (or equivalently, a dip in $d\chi_{ac}/dT$) at 220.5 K is frequency-independent. Note that the dip in $d\chi_{dc}/dT$ also occurs at exactly the same temperature. This observation provides strong experimental evidence for the existence of a thermodynamic PM - FM phase transition at T$_{c}$ = 220.5 K. The frequency-induced shift in the low-temperature peak is described well by the critical slowing down model for a cluster spin glass (CSG). According to this model, the transition to the CSG state takes place at T$_{g}$ = 7.7 K in the zero-frequency limit. The Aharony-Pytte (AP) model for a (\textit{d = 3, n = 3}) Heisenberg spin system with random magnetic anisotropy (RMA) yields the value ($\delta$ = 7/3) of the critical exponent for the critical isotherm at T$_{c}$ = 220.5 K, which agrees closely with the observed value of $\delta$ = 2.35(5). Moreover, the AP scaling equation of state correctly describes the observed crossover from the concave-downward to concave-upwards curvature in the low-field Arrott plot isotherms, as the temperature is increased through T$_{c}$. These findings clearly establish that the PM-FM transition is basically driven by RMA. 

For temperatures below $\approx$ 30 K, large enough RMA destroys long-range FM order by breaking up the infinite FM network into FM clusters of finite size and leads to the formation of a CSG state at temperatures T $\lesssim$ 8 K by promoting freezing of finite FM clusters in random orientations. Increasing strength of the single-ion magnetocrystalline anisotropy (and hence RMA) with decreasing temperature reflects an increase in the number of low-spin (LS) Co$^{3+}$ ions at the expense of that of high-spin (HS) Co$^{3+}$ ions. At intermediate temperatures (30 K $\lesssim T \lesssim$ 180 K), spin dynamics has contributions from the infinite FM network (fast relaxation governed by a single anisotropy energy barrier) and finite FM clusters (extremely slow stretched exponential relaxation due to hierarchical energy barriers).

\section*{Acknowledgements}
The authors acknowledge Kranti Sharma for AC susceptibility measurements. S.S.M. acknowledges the financial support from SERB, India, in the form of the national post doctoral fellowship (NPDF) award (PDF/2021/002137). R.S. acknowledge the financial support provided by the Ministry of Science and Technology in Taiwan under project numbers MOST-111-2124-M-001-009 \& MOST-110-2112-M-001-065-MY3 \& AS- iMATE-111-12. 


\bibliography{lsco10_Re}

\end{document}